### Sensory supplementation system

### based on electrotactile tongue biofeedback of head position

### for balance control

Nicolas VUILLERME[1,2], Nicolas PINSAULT[1], Olivier CHENU[1], Jacques DEMONGEOT[1],

Yohan PAYAN[1], Yuri DANILOV[2]

[1] Laboratoire TIMC-IMAG, UMR UJF CNRS 5525, La Tronche, France

[2] Wicab, Inc., Middleton, WI, 53562, US

18 pages (including figures); 3 figures and no table.

**Address for correspondence:**
Nicolas VUILLERME
Laboratoire TIMC-IMAG, UMR UJF CNRS 5525
Faculté de Médecine
38706 La Tronche cédex
France.
Tel: (33) (0) 4 76 63 74 86
Fax: (33) (0) 4 76 51 86 67
Email: nicolas.vuillerme@imag.fr

**Acknowledgements**
We are indebted to Professor Paul Bach-y-Rita for introducing us to the Tongue Display Unit and for discussions about sensory substitution. Paul has been for us more than a partner or a supervisor: he was a master inspiring numerous new fields of research in many domains of neurosciences, biomedical engineering and physical rehabilitation. The authors would like thank CMC Les Petites Roches, Saint Hilaire du Touvet, CHU de Grenoble to allow us using the force platform. This research was supported by Wicab Inc., Floralis (Université Joseph Fourier, Grenoble) and Fondation Garches. Special thanks also are extended to Zora B. for various contributions.

**Abstract**

The present study aimed at investigating the effects of an artificial head position-based tongue-placed electrotactile biofeedback on postural control during quiet standing under different somatosensory conditions from the support surface. Eight young healthy adults were asked to stand as immobile as possible with their eyes closed on two Firm and Foam support surface conditions executed in two conditions of No-biofeedback and Biofeedback. In the Foam condition, a 6-cm thick foam support surface was placed under the subjects' feet to alter the quality and/or quantity of somatosensory information at the plantar sole and the ankle. The underlying principle of the biofeedback consisted of providing supplementary information about the head orientation with respect to gravitational vertical through electrical stimulation of the tongue. Centre of foot pressure (CoP) displacements were recorded using a force platform. Larger CoP displacements were observed in the Foam than Firm conditions in the two conditions of No-biofeedback and Biofeedback. Interestingly, this destabilizing effect was less accentuated in the Biofeedback than No-biofeedback condition. In accordance with the sensory re-weighting hypothesis for balance control, the present findings evidence that the availability of the central nervous system to integrate an artificial head orientation information delivered through **electrical** stimulation of the tongue to limit the postural perturbation induced by alteration of somatosensory input from the support surface.

**Introduction**

      Biofeedback systems for balance control consist in supplying individuals with additional artificial information about body orientation and motion to substitute or supplement the natural visual, somatosensory and vestibular sensory cues. Among the possible alternative sensory channels that can be used to convey body-motion information, normally provided by the human senses, the somatosensory system of the tongue has recently received a growing interest [3,24,29,31,37]. Interestingly, because of its dense mechanoreceptive innervations [23] and large somatosensory cortical representation [20], the tongue can convey higher-resolution information than the skin can [22,26]. In addition, the presence of an electrolytic solution, saliva, also insures a highly efficient electrical contact between the electrodes and the tongue surface and therefore does not require high voltage and current [2]. Finally, the tongue is located in the protected environment of the mouth and is normally out of sight and out of the way, which could make a tongue-placed tactile display aesthetically acceptable.

      Following train of thought, an head position-based tongue placed biofeedback system has recently been designed to transmit artificially sensed head orientation with respect to gravitational vertical, normally provided by the vestibular system (e.g. [9]), through electrical stimulation of the tongue [3,24]. In a recent study, the effectiveness of this system in improving balance control in subjects with bilateral vestibular dysfunction has been demonstrated [24]. In the context of the multisensory control of balance (e.g., [16]), these results evidence the ability of the central nervous system (CNS) to efficiently integrate an artificial head position-based, tongue-placed electrotactile biofeedback for controlling posture, as a *sensory substitution* for loss of vestibular information. The present experiment was designed to investigate whether the CNS is able to integrate this biofeedback for balance control, as a *sensory supplementation*, to compensate for an alteration of somatosensory information, known to play a major role in postural control during quiet standing (e.g.,

[13,17,36]). To achieve this goal, we compared the effects of this artificial head position-based, tongue-placed electrotactile biofeedback [3,24] on postural control during quiet standing under different somatosensory conditions from the support surface.

**Materials and Methods**

Subjects

Eight young healthy adults (5 males and 3 females; age = 28.9 ± 7.4 years; body weight = 72.5 ± 7.2 kg; height 175.5 ± 7.7 cm; mean ± SD) with no history of motor problems, neurological disease, or vestibular impairment voluntarily participated in the experiment. They gave their informed consent to the experimental procedure as required by the Helsinki declaration (1964) and the local Ethics Committee.

Task and procedures

Eyes closed, subjects stood barefoot on a force platform with their feet performing an angle of 30° relative to each other, heels 5 cm apart and their hands loosely hanging at the sides. The force platform (Satel, Blagnac, France) allowed measuring the displacements of the centre of foot pressure (CoP). Signals from the force platform were sampled at 40 Hz (12 bit A/D conversion) and filtered with a second-order Butterworth filter with a 6-Hz low-pass cut-off frequency.

Subject's task was to sway as little as possible on two *Firm* and *Foam* support surface conditions. The force platform served as the Firm support surface. In the Foam condition, a 6-cm thick foam support surface, altering the quality and/or quantity of somatosensory information at the plantar sole and the ankle, was placed under the subjects' feet (e.g., [10,12,34,36,38]).

These two conditions were executed under two experimental sessions of No-biofeedback and Biofeedback. In the Biofeedback session, subjects performed the postural task using an head position-based tongue-placed electrotactile biofeedback (BrainPort Balance Device, Wicab Inc.) [3,6,24]. The underlying principle of the biofeedback consisted of providing supplementary information about the head orientation with respect to gravitational vertical through electrical stimulation of the tongue. In short, instantaneous pitch and roll angles of the head relative to the gravitational vertical were derived by double integration of acceleration data sensed with a micro-electromechanical system (MEMS) accelerometer and displayed on a 100-point electrotactile array held against the anterior dorsal of the tongue ($10 \times 10$ matrix of 1.5 mm diameter gold-plated electrodes on 2.32 mm centers) (Tongue Display Unit, TDU) [2]. Both the MEMS accelerometer and the electrotactile array are integrated in a custom-formed dental retainer, which subjects kept in their mouth all over the duration of the experiment (i.e. in both the No-biofeedback and Biofeedback experimental sessions). In the Biofeedback session, subjects were asked to actively and carefully hold their tongue against the matrix of electrodes that allowed them to continuously perceive both position and motion of a small "target" stimulus on the tongue display, corresponding to head orientation with respect to gravitational vertical. Specifically, as illustrated in Figure 1, when the subject's head sways on the left, right, forwards and backwards, the electrical stimulation on the tongue moves to the left, right, forward and backward, respectively. Subjects were then asked to continuously adjust head orientation and to maintain the stimulus pattern at the centre of the display [3,24]. Several practice runs were performed prior to the test to ensure that subjects had mastered the relationship between the different head positions and lingual electrical stimulations.

------------------------------------

Please insert Figure 1 about here

-------------------------------------

Three 50s trials for each condition were performed. The order of presentation of the two *Firm* and *Foam* support surface conditions and the No-biofeedback and Biofeedback experimental sessions was counterbalanced.

Data analysis

CoP displacements were processed through a space-time domain analysis including the calculation of (1) the surface area (mm²) covered by the trajectory of the CoP with a 90% confidence interval, and (2) the length of the CoP displacements (mm) along the medio-lateral (ML) and antero-posterior (AP) axes, corresponding to the sum of the displacement scalars obtained along the ML and AP axes, respectively.

Statistical analysis

Two Biofeedback (No-biofeedback *vs.* Biofeedback) × 2 Support surface (Firm *vs.* Foam) analyses of variances (ANOVAs) with repeated measures of both factors were applied to data. Post hoc analyses (*Newman-Keuls*) were performed whenever necessary. Level of significance was set at 0.05.

**Results**

Figure 2 illustrates representative displacements of the CoP from a typical subject during standing in each of the four experimental conditions: No-biofeedback / Firm (2A) No-biofeedback / Foam (2B), Biofeedback / Firm (2C) and Biofeedback / Foam (2D).

-------------------------------------

Please insert Figure 2 about here

-------------------------------------

Analysis of the surface area covered by the trajectory of the CoP showed a significant interaction of Support surface × Biofeedback condition ($F(1,7) = 16.70$, $P < 0.01$). As illustrated in Figure 3A, the decomposition of this interaction into its simple main effects indicated a larger stabilizing effect of Biofeedback on the Foam ($P < 0.001$) than Firm condition ($P < 0.05$). The ANOVAs also showed main effects of Support surface ($F(1,7) = 260.43$, $P < 0.001$) and Biofeedback ($F(1,7) = 44.74$, $P < 0.001$), yielding increased surface area in the Foam relative to the Firm condition and decreased surface area in the Biofeedback relative to the No-biofeedback condition, respectively.

Analyses of the length of the CoP displacements along both the ML and AP axes showed significant interactions of Support surface × Biofeedback condition ($F(1,7) = 6.71$, $P < 0.05$ and ($F(1,7) = 10.43$, $P < 0.05$, for ML and AP axes, respectively). As illustrated in Figures 3B and 3C, the decomposition of this interaction into its simple main effects indicated a larger stabilizing effect of Biofeedback in the Foam ($Ps < 0.001$) than Firm condition ($Ps < 0.05$). The ANOVAs also showed main effects of Support surface ($F(1,7) = 164.82$, $P < 0.001$ and $F(1,7) = 207.76$, $P < 0.001$, for ML and AP axes, respectively) and Biofeedback ($F(1,7) = 36.71$, $P < 0.001$ and $F(1,7) = 15.38$, $P < 0.01$, for ML and AP axes, respectively), yielding increased length of the CoP displacements in the Foam relative to the Firm condition and decreased length of the CoP displacements in the Biofeedback relative to the No-biofeedback condition, respectively.

------------------------------------

Please insert Figure 3 about here

------------------------------------

**Discussion**

The present study aimed at investigating the effects of an artificial head position-based tongue-placed electrotactile biofeedback on postural control during quiet standing under different somatosensory conditions from the support surface. To achieve this goal, eight young healthy adults were asked to stand as immobile as possible with their eyes closed on two Firm and Foam support surface conditions executed during two No-biofeedback and Biofeedback experimental sessions. In the Foam condition, a 6-cm thick foam support surface was placed under the subjects' feet to alter the quality and/or quantity of somatosensory information at the plantar sole and the ankle. The underlying principle of the biofeedback consisted of providing supplementary information about the head orientation with respect to gravitational vertical through electrical stimulation of the tongue (Fig. 1). Note that all subjects were able to complete the test without reporting any pain or discomfort. Centre of foot pressure (CoP) displacements were recorded using a force platform.

On the one hand, standing on a compliant foam surface deteriorated postural control, as indicated by the increased surface area (Fig. 3A) and length of the CoP displacements along the ML (Fig. 3B) and AP (Fig. 3C) axes observed in the Foam relative to the Firm condition. This result corroborate previous observations (e.g., [10,12,34,36,38]). Together with the postural effects previously observed when anaesthetising (e.g. [17]), cooling (e.g., [1]) or stimulating (e.g., [4,15,21,25]) the plantar soles, i.e., when manipulating somatosensory information from plantar cutaneous receptors, these results add to the large body of evidence suggesting the importance of somatosensory inputs from the plantar soles and ankles in postural control during quiet standing (e.g., [13,17,36]).

On the other hand, the availability of the biofeedback improved postural control, as indicated by the decreased surface area (Fig. 3A) and length of the CoP displacements along the ML (Fig. 3B) and AP (Fig. 3C) axes observed in the Biofeedback relative to the No-biofeedback condition. This result confirms the ability of the CNS to integrate an artificial head orientation information delivered through electrical stimulation of the tongue to improve postural control [3,24]. Note that the TDU already has proven its efficiency when used as the sensory output unit for visual [22], tactile [29,31,37] and proprioceptive [27,30] substitution or augmentation applications.

More originally, the availability of the biofeedback allowed the subjects to limit the destabilizing effect induced by the alteration of somatosensory input from the support surface, as indicated by the significant interactions Support surface × Biofeedback observed for the surface area (Fig. 3A) and the length of the CoP displacements along the ML (Fig. 3B) and AP (Fig. 3C) axes. These results could be attributable to the sensory re-weighting hypothesis (e.g. [18,19, 28,32,34,36,38]), whereby the CNS dynamically and selectively adjusts the relative contributions of sensory inputs (i.e., the sensory weights) to maintain upright stance depending on the sensory contexts. For instance, in the condition of ankle muscle fatigue, known to alter proprioceptive signals from the ankle [27], the sensory integration process involved in the control of bipedal postural control has been shown to (1) decrease the contribution of proprioceptive cues from the ankle [32], and (2) increase the contribution of vision [14,28], cutaneous inputs from the foot and shank [33] and haptic cues from the finger [35], providing reliable and accurate sensory information for controlling posture. In the present experiment, the decreased CoP displacements observed in the Foam condition when the Biofeedback was in use relative to when it was not, suggests an increased reliance on sensory information related to the head orientation with respect to gravitational vertical, i.e. closely related to vestibular inputs (e.g., [9]), in condition of altered somatosensory information from the support surface. Note that these results are consistent with the increased postural responses to vestibular perturbation previously observed when somatosensory information from the support surface was altered either in healthy subjects by standing on a compliant (e.g. [10]), on a sway-referenced (e.g. [5]), unstable (e.g. [8]) or moving support surface (e.g. [11]), or by somatosensory loss due to neuropathy (e.g. [7,10]).

Finally, in addition to their fundamental relevance on the field of neuroscience, we believe that the present findings could complementarily have implications in clinical conditions and rehabilitation practice. With this context, we are presently exploring whether head-position information, when presented to the tongue via electrical stimulation, could positively affect postural control in individuals with somatosensory loss in the feet from diabetic peripheral neuropathy and persons with lower limb amputation.

**References**

[1]    H. Asai, K. Fujiwara, H. Toyoma, T. Yamashina, I. Nara, K. Tachino, The influence of foot soles cooling on standing postural control, in: T.H. Brandt, W. Paulus, W. Bles, M. Deitrich, S. Krafezyk, A. Straube (Eds.), Disorders of Posture and Gait, Thieme, Stuttgart, 1990, pp. 198-201.

[2]    P. Bach-y-Rita, K.A. Kaczmarek, M.E. Tyler, J. Garcia-Lara, Form perception with a 49-point electrotactile stimulus array on the tongue. J. Rehabil. Res. Dev. 35 (1998) 427-430.

[3]    P. Bach-y-Rita, S.W. Kercel, Sensory substitution and the human-machine interface. Trends Cogn. Sci. 7 (2003) 541-546.

[4]    L. Bernard-Demanze, N. Vuillerme, L. Berger, P. Rougier, Magnitude and duration of the effects of plantar sole massages. Int. SportMed J. 7 (2006) 154-169.

[5]    M. Cenciarini, R.J. Peterka, Stimulus-dependent changes in the vestibular contribution to human postural control. J. Neurophysiol. 95 (2006) 2733-2750.

[6]    Y. Danilov, M. Tyler, Brainport: an alternative input to the brain. J. Integr. Neurosci. 4 (2005) 537-550.

[7]    B.L. Day, J. Cole, Vestibular-evoked postural responses in the absence of somatosensory information. Brain 125 (2002) 2081-2088.

[8]    R. Fitzpatrick, D. Burke D, S.C. Gandevia, Task-dependent reflex responses and movement illusions evoked by galvanic vestibular stimulation in standing humans. J. Physiol. Lond. 478 (1994) 363-372.

[9]    A.M. Green, D.E. Angelaki, An integrative neural network for detecting inertial motion and head orientation. J. Neurophysiol. 92 (2004) 905-925.

[10]   F.B. Horak, F. Hlavacka, Somatosensory loss increases vestibulospinal sensitivity. J. Neurophysiol. 86 (2001) 575-585.

[11] J.T. Inglis, C.L. Shupert, F. Hlavacka, F.B. Horak, The effect of galvanic vestibular stimulation on human postural responses during support surface translations. J. Neurophysiol. 73 (1995) 896-901.

[12] B. Isableu, N. Vuillerme, Differential integration of kinesthetic signals to postural control. Exp. Brain Res. 174 (2006) 763-768.

[13] A. Kavounoudias, R. Roll, J.P. Roll, The plantar sole is a "dynamometric map" for human balance control. NeuroReport 9 (1998) 3247-3252.

[14] T. Ledin, P.A. Fransson, M. Magnusson, Effects of postural disturbances with fatigued triceps surae muscles or with 20% additional body weight. Gait & Posture 19 (2004) 184-193.

[15] B.E. Maki, S.D. Perry, R.G. Norrie, W.E. McIlroy, Effect of facilitation of sensation from plantar foot-surface boundaries on postural stabilization in young and older adults. J. Gerontol. A Biol. Sci. Med. Sci. 54 (1999) M281-M287.

[16] J. Massion, Postural control system. Curr. Opin. Neurobiol. 4 (1994) 877-887.

[17] P.F. Meyer, L.I. Oddsson, C.J. De Luca, The role of plantar cutaneous sensation in unperturbed stance. Exp. Brain Res. 156 (2004) 505-512.

[18] R.J. Peterka, Sensorimotor integration in human postural control. J. Neurophysiol. 88 (2002) 1097-1118.

[19] R.J. Peterka, P.J. Loughlin, Dynamic regulation of sensorimotor integration in human postural control. J. Neurophysiol. 91 (2004) 410-423.

[20] C. Picard, A. Olivier, Sensory cortical tongue representation in man. J. Neurosurg. 59 (1983) 781-789.

[21] A. Priplata, J. Niemi, M. Salen, J. Harry, L.A. Lipsitz, J.J. Collins, Vibrating insoles and balance control in elderly people. Lancet 362 (2003) 1123-1124.

[22] E. Sampaio, S. Maris, P. Bach-y-Rita, Brain plasticity: 'visual' acuity of blind persons via the tongue. Brain Res. 908 (2001) 204-207.

[23] M. Trulsson, G.K. Essick, Low-threshold mechanoreceptive afferents in the human lingual nerve. J. Neurophysiol. 77 (1997) 737-748.

[24] M. Tyler, Y. Danilov, P. Bach-y-Rita, Closing an open-loop control system: vestibular substitution through the tongue. J. Integr. Neurosci. 2 (2003) 159-164.

[25] J. Vaillant, N. Vuillerme, A. Janvy, F. Louis, R. Braujou, R. Juvin, V. Nougier, Effect of manipulation of the feet and ankles on postural control in elderly adults. Brain Res. Bull. (2007) (doi:10.1016/j.brainresbull.2007.07.009).

[26] R.W. van Boven, K.O. Johnson, The limit of tactile spatial resolution in humans: grating orientation discrimination at the lips, tongue, and finger. Neurology 44 (1994) 2361-2366.

[27] N. Vuillerme, M. Boisgontier, O. Chenu, J. Demongeot, Y. Payan, Tongue-placed tactile biofeedback suppresses the deleterious effects of muscle fatigue on joint position sense at the ankle. Exp. Brain Res. 183 (2007) 235-240.

[28] N. Vuillerme, C. Burdet, B. Isableu, S. Demetz, The magnitude of the effect of calf muscles fatigue on postural control during bipedal quiet standing with vision depends on the eye-visual target distance. Gait & Posture 24 (2006) 166-172.

[29] N. Vuillerme, O. Chenu, J. Demongeot, Y. Payan, Controlling posture using a plantar pressure-based, tongue-placed tactile biofeedback system. Exp. Brain Res. 179 (2007) 409-414.

[30] N. Vuillerme, O. Chenu, J. Demongeot, Y. Payan, Improving human ankle joint position sense using an artificial tongue-placed tactile biofeedback. Neurosci. Lett. 405 (2006) 19-23.

[31] N. Vuillerme, O. Chenu, N. Pinsault, M. Boisgontier, J. Demongeot, Y. Payan, Inter-individual variability in sensory weighting of a plantar pressure-based, tongue-placed tactile biofeedback for controlling posture. Neurosci. Lett. 421 (2007) 173-177.

[32] N. Vuillerme, F. Danion, N. Forestier, V. Nougier, Postural sway under muscle vibration and muscle fatigue in humans. Neurosci. Lett. 333 (2002) 131-135.

[33] N. Vuillerme, S. Demetz, Do ankle foot orthoses modify postural control during bipedal quiet standing following a localized fatigue at the ankle muscles? Int. J. Sports Med. 28 (2007) 243-246.

[34] N. Vuillerme, L. Marin, B. Debû, Assessment of static postural control in teenagers with Down Syndrome. Adapt Phys Act Q 18 (2001) 417-433.

[35] N. Vuillerme, V. Nougier, Effect of light finger touch on postural sway after lower-limb muscular fatigue. Arch. Phys. Med. Rehabil. 84 (2003) 1560-1563.

[36] N. Vuillerme, N. Pinsault, Re-weighting of somatosensory inputs from the foot and the ankle for controlling posture during quiet standing following trunk extensor muscles fatigue. Exp. Brain Res. (2007) (doi 10.1007/s00221-007-1047-4.).

[37] N. Vuillerme, N. Pinsault, O. Chenu, M. Boisgontier, J. Demongeot, Y. Payan, How a plantar pressure-based, tongue-placed tactile biofeedback modifies postural control mechanisms during quiet standing. Exp. Brain Res. 181 (2007) 547-554.

[38] N. Vuillerme, N. Pinsault, J. Vaillant, Postural control during quiet standing following cervical muscular fatigue: effects of changes in sensory inputs. Neurosci. Lett. 378 (2005) 135-136.

**Figure captions**

**Figure 1.** Sensory coding schemes for the Tongue Display Unit (TDU) (*right panel*) as a function of the head orientation with respect to gravitational vertical (*left panel*). (1) Neutral, (2) right-side-bended, (3) left-side-bended, (4) flexed and (5) extended head postures.

**Figure 2.** Representative displacements of the centre of foot pressure (CoP) from typical subjects during standing in each of the four experimental conditions: No-biofeedback / Firm (A), No-biofeedback / Foam (B), Biofeedback / Firm (C) and Biofeedback / Foam (D).

**Figure 3.** Mean and standard deviation of the surface area (A) and the length of the CoP displacements along the medio-lateral (B) and antero-posterior (C) axes obtained in the two conditions of Firm and Foam and the two conditions of No-biofeedback and Biofeedback. The two conditions of No-biofeedback and Biofeedback are presented with different symbols: No-biofeedback (*white bars*) and Biofeedback (*black bars*). The significant *P*-values for comparisons between the No-biofeedback and Biofeedback conditions also are reported (*: $P < 0.05$; ***: $P < 0.001$).

**Figure 1**

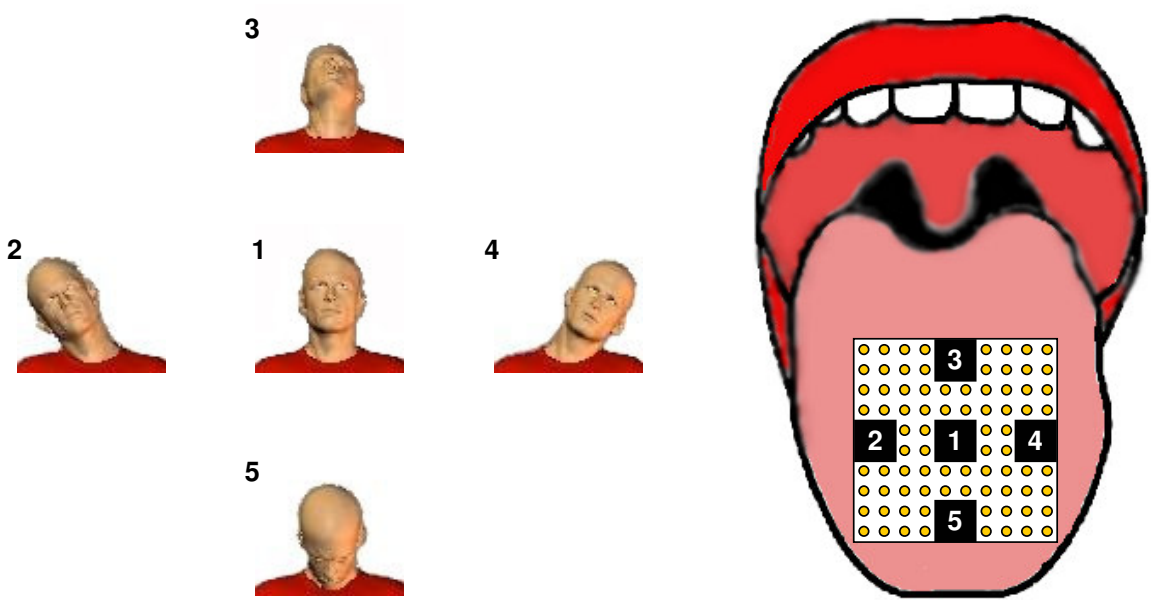

**Figure 2**

**A. No-biofeedback / Firm**     **B. No-biofeedback / Foam**

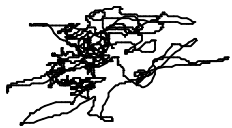 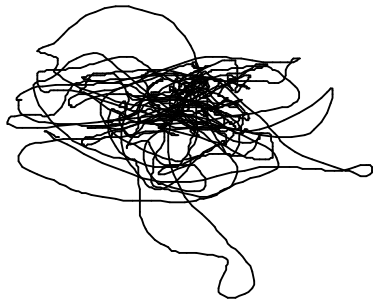

**C. Biofeedback / Firm**     **D. Biofeedback / Foam**

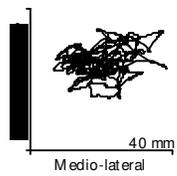 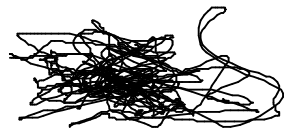

40 mm

Medio-lateral

**Figure 3**

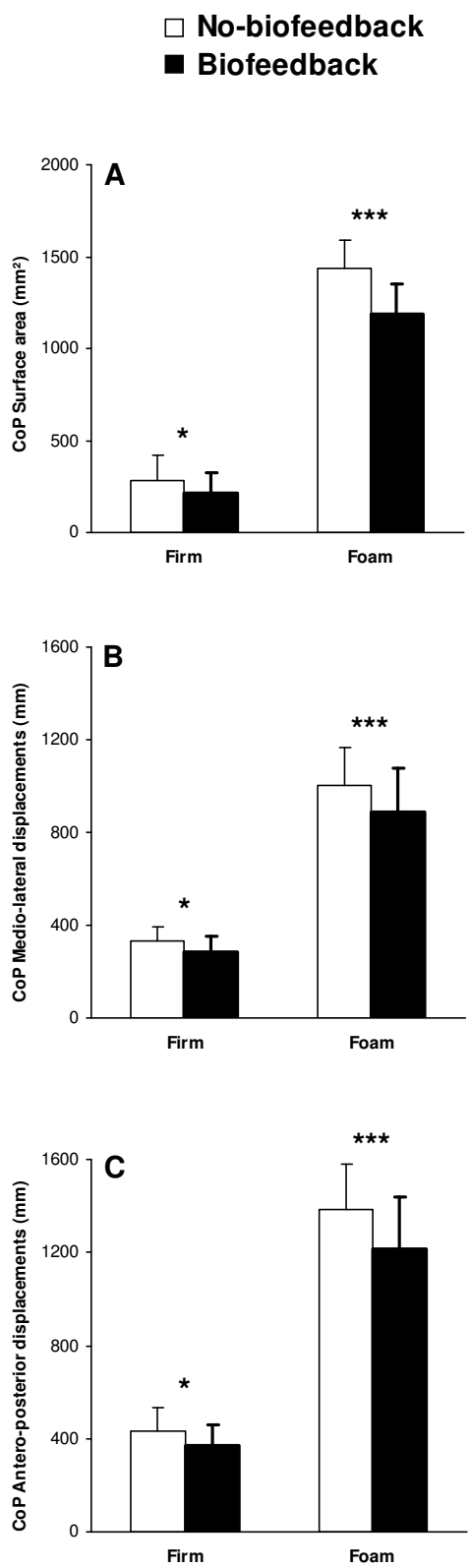